\documentstyle[11pt,aaspp4]{article}
\received{}
\accepted{}
\journalid{}{}

\slugcomment{}
\lefthead{Korchagin, Tsuchiya, Miyama}
\righthead{}
\newcommand{\be}{\begin{equation}}
\newcommand{\ee}{\end{equation}}
\begin{document}

\title{On the Origin of Faint Intracluster Starlight in Coma}
\author{Vladimir Korchagin}
\affil{Institute of Physics, Stachki 194, Rostov-on-Don 344090, Russia\\
Email: vik@rsuss1.rnd.runnet.ru}
\author{Toshio Tsuchiya\altaffilmark{1}}
\affil{Department of Astronomy, Faculty of Science, Kyoto University,
  Japan \\
Department of Earth and Space Science, Osaka University, Japan\\
Email: tsuchiya@kusastro.kyoto-u.ac.jp }
\altaffiltext{1}{Current Affiliation: Astronomisches Rechen-Institut,
  Heidelberg D-69120, Germany}
\and
\author{Shoken M. Miyama}
\affil{National Astronomical Observatory, Mitaka, Tokyo 181, Japan\\
     Email: miyama@th.nao.ac.jp}
\begin{abstract}
Using N-body numerical simulations, we examine possible
mechanisms for the origin of the intracluster starlight features
recently discovered in the Coma cluster of galaxies.
We show that tidal interactions of a ``normal'' elliptical galaxy
with a symmetrically distributed dark matter potential
do not produce the observed intracluster starlight features.
A head-on collision of two normal ellipticals does, however, explain the origin
of the intracluster features with surface brightness about 26 mag/arcsec$^2$). Another possible explanation for the
intracluster starlight features is
galactic tidal interactions
with massive intracluster substructures.
The presence of substructures in Coma with a density of
$\sim$ 0.1 times the density in the center of the galaxy,
and with a total mass about three orders of magnitude larger than
the mass of a normal elliptical galaxy, would account for the
observations.
\end{abstract}
%
%
      \section{INTRODUCTION}
%

Gregg \& West (1998) have recently reported the discovery of 
intracluster starlight features in the Coma cluster of galaxies. 
The features are revealed in R-band images of the Coma cluster, and are also
seen in the B, V and I bands, confirming their existence.
The plume-like features are tens of kiloparsecs in width,
and extend to radii of more than a hundred kiloparsecs.  
The mean R-band surface brightness of the plumes is low,
about 26 mag/arcsec$^2$, but their absolute magnitude 
$\approx - 19.0$ mag is comparable to the 
absolute R-band magnitude of a normal elliptical galaxy.
Possible explanations of the origin of the plumes
invoke the idea of the disruption
of galaxies by galaxy-galaxy interactions (Dubinski et al 1996, Moore et al 1996),
or the tidal stripping of galaxies by the global cluster gravitational
field (Merritt 1984).
However, the stellar plumes observed in the Coma cluster can not be produced so easily.
Using N-body simulations, we have studied the possibility of forming
stellar plumes with
surface brightnesses 6--7 magnitudes fainter than the surface
brightness in the centers of the normal elliptical galaxies similar
to those observed in the Coma cluster.
We have found that the
tidal interaction of galaxies
with the gravitational field of the cluster core cannot be responsible
for the origin of intracluster starlight features in Coma.
A head-on collision of two normal elliptical galaxies
with a relative velocity less than the cluster velocity dispersion, 
or tidal interactions with intracluster density inhomogeneities
might explain the origin of stellar plumes with the observed properties.

%
        \section{MODELS AND NUMERICAL METHOD}

To model galaxy disruptions by
the cluster core or galaxy-galaxy collisions, we use the
parameters of the elliptical galaxy NGC 3377.
The disruptions of the disk galaxies might also lead to the formation
of stellar plumes. 
Nearby clusters of galaxies, however, are enriched with
elliptical and lenticular galaxies (Dressler 1980).
The elliptical galaxy NGC 3377 was chosen by Gregg \& West (1998)
in their comparison between normal galaxies and the
properties of the Coma stellar plumes.
NGC 3377 is well studied observationally.
Kormendy et al (1998) published an accurate rotation
curve of this galaxy, along with a
composite surface brightness profile along its major axis.
The rotation curve of NGC 3377 shows a steep kinematic gradient near the
center of the galaxy, but it remains flat for radii $r > 3^{\prime \prime}$,
with a circular velocity about 100 km/sec.
This yields a total mass for NGC 3377 of
about $3\times 10^{10} M_{\odot}$, if the galactic radius is
$\sim 140^{\prime \prime}$ or 15 kpc assuming a Hubble constant 
$H = 50 \mathrm{km s}^{-1} \mathrm{Mpc}^{-1}$, as used in our further
simulations. 

Using the rotational curve near the center, the central density
of NGC 3377 can be rather firmly evaluated to be
$2.2 \times 10^{-21}$ g/cm$^3$
within the central $3^{\prime \prime}$.
Approximately half of the mass can be
associated with the central black hole, and we adopt in our simulations
the value $10^{-21}$ g/cm$^3$.

For the radial dependence of the
galactic density, we use the Plummer model,
and Hernquist's (1990) density distributions with 
core radii of 0.8 kpc and 0.6 kpc respectively. 
We adopt an analytic King approximation to the isothermal sphere (Sarazin 1986)
for the dark matter density distribution in the Coma cluster. 
Hughes (1998) re-examined mass estimates for the Coma cluster derived from
X-ray spectral observations. At the 99\% confidence level, he found that the
mass of the Coma cluster within a radius of 1 Mpc is 
$(6.2 \pm 0.9) \times 10^{14} M_{\odot}$. The dark matter core radius
for the adopted matter profile is about
270-370 kpc (Hughes 1998), which gives a central dark matter density 
of $(1.0 - 2.9)\times 10^{15} M_{\odot}$ / Mpc$^3$.  
In our simulations, we adopt ``extreme'' disruptive parameters
of the Coma cluster which have the values
$r_c = 270$ kpc, and $\rho_{0} = 3\times 10^{15} M_{\odot}$ / Mpc$^3$.

The simulations were performed with a TREE code using 
N=32768 particles,  a tolerance parameter equal to 0.75, and a softening length
of 0.03125. 

%
%
        \section{RESULTS AND DISCUSSION}

We considered different possibilities for the formation of stellar plumes
in clusters of galaxies, namely the tidal disruption of an elliptical galaxy
by the cluster core, direct galactic collisions, and the
interaction of an elliptical galaxy with hypothetical cluster substructures. 

Collisions or tidal interactions often produce plumes of particles in
numerical simulations. We need, however, a quantitative estimate of
whether such plumes could be responsible for the observed stellar
structures seen in Coma. To do this, we calculated the theoretical
surface brightness in the plume assuming constant mass-to-luminosity
ratios for the plume and the galaxy.  The characteristic surface
brightness in the central regions of elliptical galaxies in the V-band
is about 20 mag/arcsec$^2$ (Binney \& Merrifield 1998), which for the
R-band, gives a value of about 19.5 mag/arcsec$^2$ (Frei \& Gunn 1994).
We assumed that our model galaxy has the same surface brightness in
central regions before an interaction, and that tidal interactions do
not change the stellar content of the plume. These assumptions allow us
to estimate surface brightness of resulting features in our numerical
simulation by comparing corespondinc column densities of particles. The
last assumption excludes from our consideration the possible role of
star formation triggered in the plumes by tidal interactions.  Although
tidally induced star formation in the plumes is likely to take place for
the collisions of gas rich disk galaxies, this is not the case for the
stellar plumes observed in Coma. According to Gregg \& West (1998), the
colors in the plumes are consistent with those of evolved stellar
polulations observed in intermediate brightness normal elliptical
galaxies.  The photometric uncertainties, and uncertainties of the
removal of superimposed objects leave, however, open the possibility of
significant star fromation in the plumes.

{\it Direct galactic collisions}. Figure 1 shows the result 
of a direct collision of two equal-mass Plummer spheres 
taken at time $ 1.7\times 10^8$ years after the beginning of the interaction.
The initial separation of the
galaxies and their initial relative velocity were taken to be
32 kpc and 408 km/sec.
The collision forms a broad stellar plume connecting both galaxies, with the
surface brightness in the plume about 26.5 mag/arcsec$^2$(Figure 2).
The direct collisions of equal-mass galaxies
might therefore constitute a possible explanation of the
origin of the observed stellar plumes in Coma. The high velocity dispersion
of about 1000 km/sec in Coma, however, makes
this mechanism rather ineffective. 
Gregg \& West (1998) argue that the galactic subgroups in Coma,
with their relatively low internal velocity dispersions, will
increase the effectiveness of the disruption of galaxies,
and our simulations concur with this conclusion. 
The characteristic velocity dispersion
within groups, which are fairly uniformly distributed in space and which show
no preference for rich clusters, is about 250 km/sec (Hickson 1997).
This value is much smaller 
than the typical velocity dispersions in rich clusters of galaxies.
If a cluster is not virialized, then strong local
interactions can occur within in-falling subgroups of galaxies
(Conselice \& Gallaher 1998). Burns et al (1994) present convincing arguments
that such infalling groups of galaxies do indeed exist.
They found that the galactic group NGC 4839 has fallen into the Coma Cluster,
passing through the cluster core about 2 Gyr ago. 
The NGC 4839 galactic group cannot be responsible, however, for the observed 
starlight features in Coma. All four stellar features are located within
$ \sim 0.5$ Mpc of the cluster center, and cannot survive for more than 1-2 Gyr 
(Gregg \& West 1998),
which is shorter than the epoch of passage of the NGC 4839 group
through the central regions of Coma.

{\it Core tidal interactions}. The tidal interaction of galaxies with the 
gravitational potential of the cluster core
is a likely explanation for the origin
of stellar intracluster structures in Coma. 
We simulated an extreme case of such an interaction
with a galaxy plunging into the cluster core. 
In these simulations, the cluster core radius adopted was 270 kpc, 
and the core density equal to $3\times 10^{15} M_{\odot}$ / Mpc$^3$
which are the extreme ``disruptive'' parameters of the mass distribution in Coma.  
The galaxy was placed 600 kpc away from the cluster center with zero initial 
relative velocity.  
Figure 3 shows surface density contour plots taken at time $T=3.36 \times 10^9$ 
years after the beginning of the simulations, when the galaxy made one
oscillation in the cluster core.
The surface brightness in the patchy plume is 32-34 mag/arcsec$^2$, and apparently cannot
explain the origin of the bright starlight features in Coma. This result is
also illustrated by Figure 4, which shows the fraction of escaped particles
as a function of time. A total of 10 \% of the mass is stripped during the
interaction time. This stripped mass cannot be regarded as the main source of
intracluster stellar material in Coma. 

{\it Tidal disruption by the ``local core''}. 
One more possibility exists for the formation of the bright intracluster structures
seen in Coma. The X-ray surface brightness measurements of Coma cluster
reveal the regions of the enhanced X-ray emission with characteristic
angular size less than the core radius of the cluster 
( Soltan \& Fabricant 1990).  
The intracluster stellar features in Coma may result
from tidal interactions with small scale density fluctuations associated
with the enhanced X-ray emission.
Vikhlinin et al (1994) found indeed
such mass concentrations 
in the core region of the Coma cluster. Both of the discovered density enhancements 
are centered at the positions of the brightest elliptical galaxies
in the cluster: NGC 4874 and NGC 4889. Assuming that X-ray surface brightness
enhancements result from the density enhancements of the underlying dark matter,
Vikhlinin et al (1994) estimated the masses of inhomogeneities within 80 kpc  
to be about $10^{13} M_{\odot}$. Remarkably, the most prominent plume-like
structure, the LSB object 1 discovered by Gregg and West, is adjacent to NGC 4874.
The core radius of the density enhancement centered at the position of NGC 4874
is about 100 kpc (Vikhlinin et al 1994) which gives the density of the 
``local core''about $5 \times 10^{-25}$ g/cm$^3$.
This value is too low to disrupt noticeably a galaxy by tidal interaction.

The parameters of the density enhancement centered at the position of NGC 4889
are more favorable for the galactic disruption.  This density enhancement has core 
radius about 24 kpc, and the central density about $8 \times 10^{-24}$ g/cm$^3$.
We simulated an interaction of a galaxy with the NGC 4889 density enhancement.
A galaxy was placed 200 kpc away from the density enhancement with the initial
velocity 500 km/s. The galaxy was allowed then to collide 
non-centrally with the density enhancement with
the impact parameter equal to 24 kpc. 
Figure 5 shows the result of such high-speed tidal interaction.
Interaction with NGC 4889 density inhomogeneity forms a 
plume with the surface brightness only 29-30 mag/arcsec$^2$.
This value is too low to account for the properties of the observed
plumes. The density enhancements, however increase an effective cross-section
of the direct collision with  the giant cD galaxies nested in their central regions.
This is a likely scenario of the formation
of the most luminous plume LSB no 1 adjusted to NGC 4884.

Stars in the plume structures which are induced in direct galactic 
collisions have higher velocity dispersion
compared to the plumes produced in tidal interactions. The collisional
plumes live therefore shorter than tidally induced structures.
Our numerical simulations show that the density inhomogeneities
tidally disrupt an elliptical galaxy if the central density of a 
``local core'' is about 10 percent of the central galactic density.
Such value can be achieved if an inhomogeneity has total mass
about one thousand of a mass of a normal elliptical galaxy, and a core
radius $\le$ 15 kpc. 
If such inhomogeneities of the intracluster matter do really
exist, 
then the starlight features in clusters of galaxies give an opportunity to probe
the cluster density fluctuations. 

%
\section{SUMMARY}

Our results can be summarized as follows:

1. We found that the observed intracluster starlight structures in Coma
cannot be produced by the tidal interaction of an elliptical
galaxy with the gravitational potential of the cluster core.

2. The head-on collision of two normal ellipticals with relative
velocities lower than the velocity dispersion of the Coma cluster
might be a mechanism for the formation of the stellar plumes observed in Coma.
The subgroups of galaxies within the Coma cluster have relatively low
velocity dispersions, and therefore direct galactic collisions
within the subgroups are currently the most plausible explanation
for the origin of the stellar plumes.

3. The stellar plumes seen in Coma may result from the tidal interactions
of galaxies with local inhomogeneities of the cluster density distribution.
If the existence of such inhomogeneities is confirmed in future observations,
then intracluster starlight features will provide an opportunity for 
studying their nature.
%

The authors thank W. van Altena, G. Laughlin and C. Boily for comments
on the manuscript resulted in a much improved presentation.  VK is
grateful to the National Astronomical Observatory Japan for the
financial support.

\newpage
%
%

\newpage

\clearpage
\newpage
\begin{figure}[htbp]
  \begin{center}
    \plotone{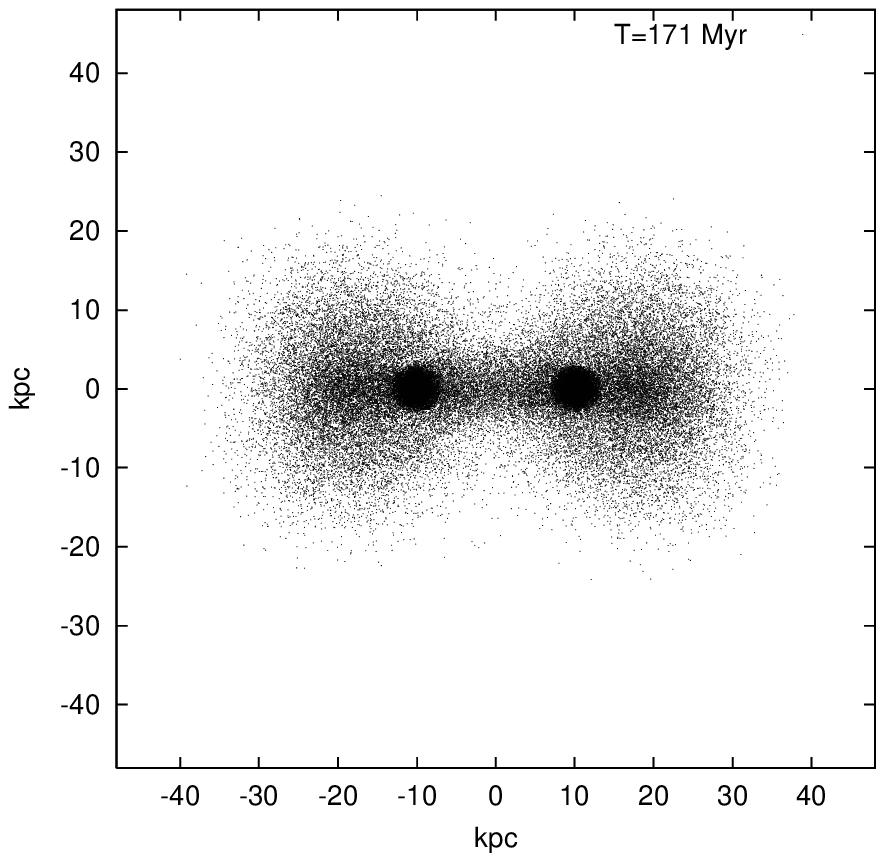}
    \figcaption[fig1.eps]{Snapshot of the collisional disruption of two
    equal mass Plummer spheres taken at time $1.7 \times 10^8$ year. The
    collision forms a broad expanding plume.}  
    \label{fig:fig1}
  \end{center}
\end{figure}
\begin{figure}[htbp]
  \begin{center}
    \plotone{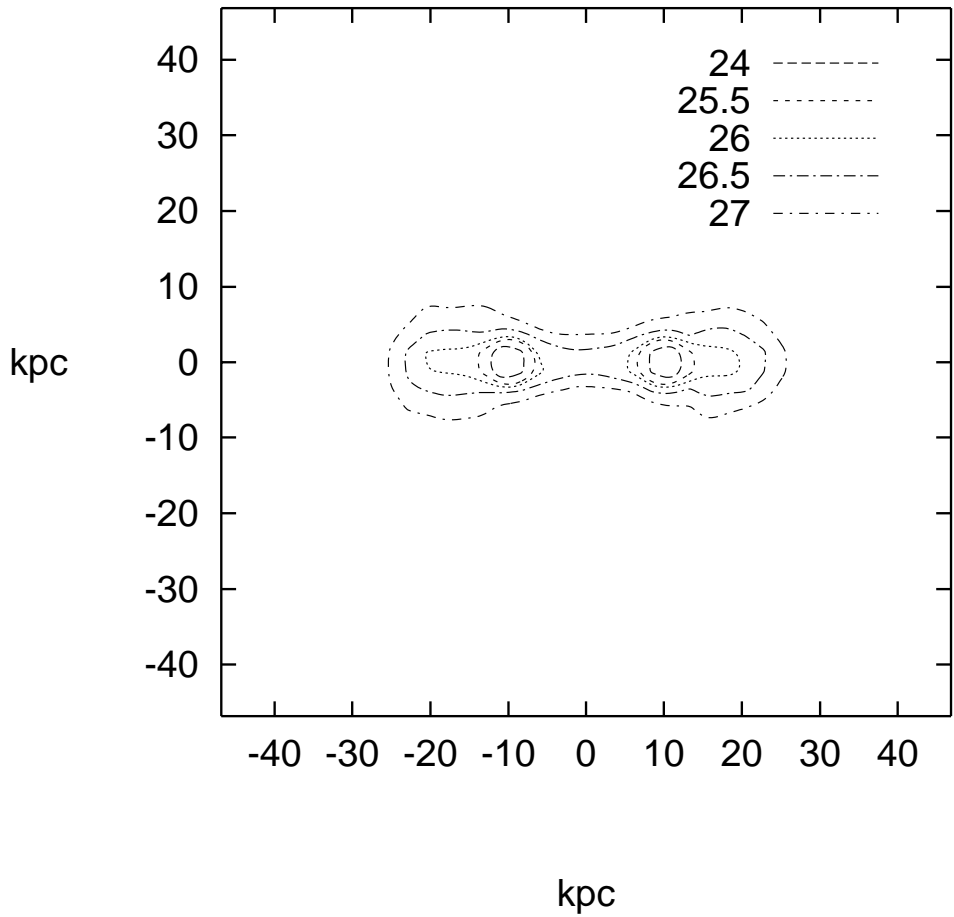}
    \figcaption[fig2.eps]{R-band surface brightness theoretical
    isophotes in mag/arcsec${}^2$ of a plume formed in the direct
    collision of two elliptical galaxies assuming their central surface
    brightness of 19.5 mag/arcsec${}^2$. The stripped particles form a
    plume of surface brightness equal to 26.5 mag/arcsec$^2$. } 
    \label{fig:fig2}
  \end{center}
\end{figure}

\begin{figure}[htbp]
  \begin{center}
    \plotone{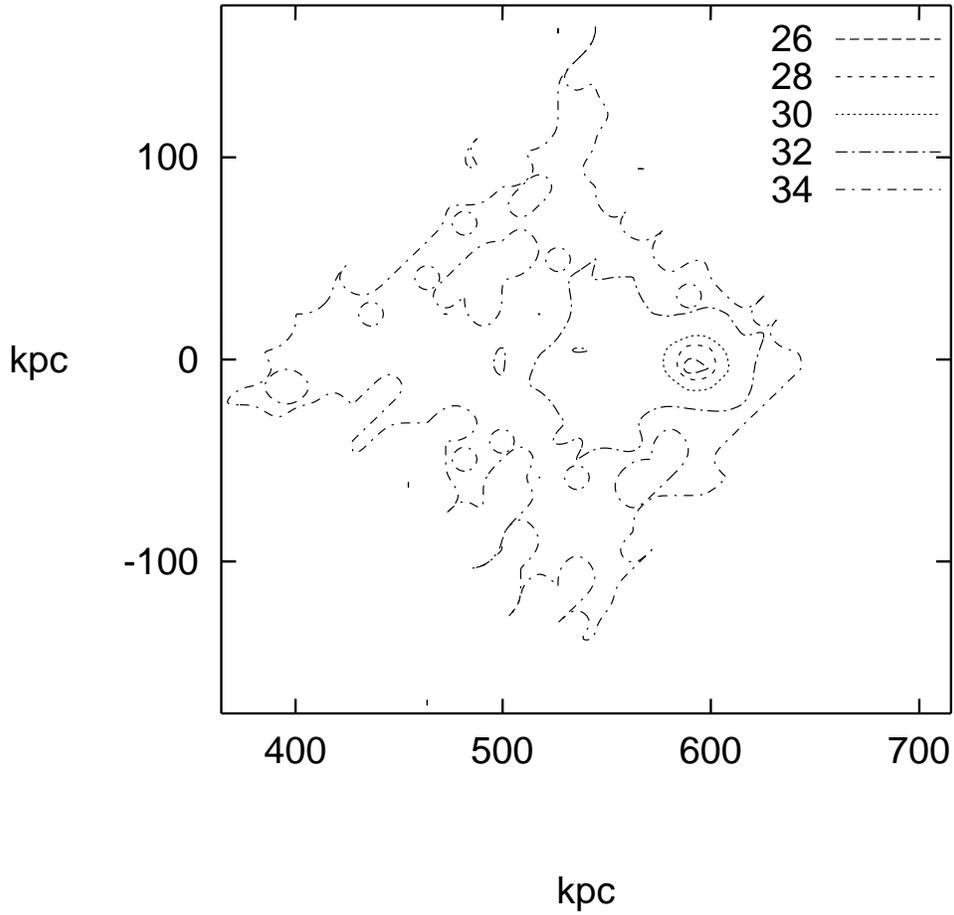}
    \figcaption[fig3.eps]{ The surface brightness isophotes in
    mag/arcsec${}^2$ for the tidal interaction of a galaxy with the
    gravitational potential of the cluster core taken at time $T=3.36
    \times 10^9$ years.  The cluster core radius is equal to 270 kpc,
    and the core density is equal to $3 \times 10^{15} M_{\odot}$ /
    Mpc$^3$. An interaction forms a broad patchy plume of the surface
    brightness equal to 32-24 mag/arcsec$^2$}
    \label{fig:fig3}
  \end{center}
\end{figure}
\begin{figure}
  \begin{center}
    \plotone{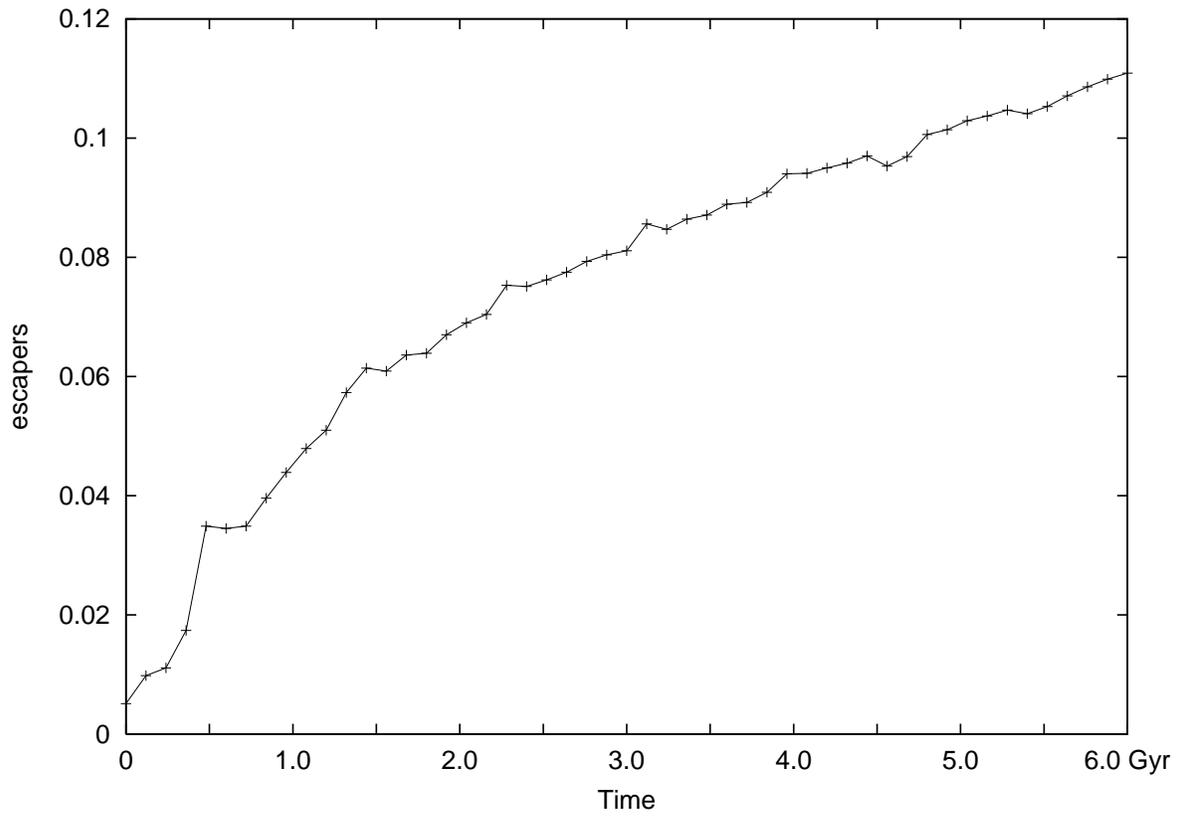}
    \figcaption[fig4.eps]{ The fraction of escaped particles as a
    function of time for the galaxy plunging into the cluster core.}
    \label{fig:fig4}
  \end{center}
\end{figure}
\begin{figure}[htbp]
  \begin{center}
    \plotone{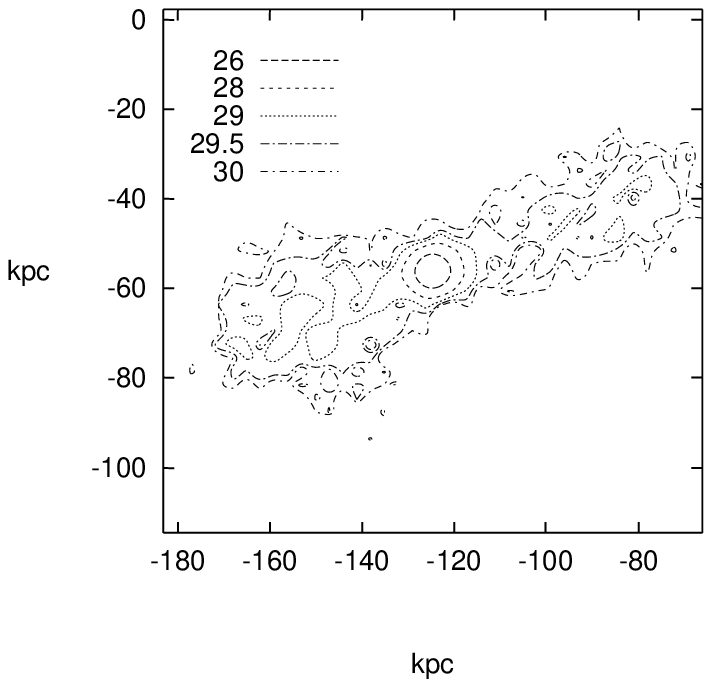}
    \caption{}
    \figcaption[fig5.eps]{ The surface density contour plots for a
      galaxy interacting with the ``local core'' inhomogeneity
      surrounding the cD elliptical galaxy NGC 4889. The core density of
      the local inhomogeneity is equal to $8 \times 10^{-24}$ g/cm$^3$;
      its core radius is equal to 24 kpc. An interaction of an
      elliptical with the ``local core'' forms a plume with a surface
      brightness of 29 - 30 mag/arcsec$^2$. }
    \label{fig:fig5}
  \end{center}
\end{figure}


\begin{thebibliography}{}

\bibitem{1} 
Binney, J., \& Merrifield, M. 1998, Galactic Astronomy,
Princeton University Press

\bibitem{3}
Burns, J.O., Roettiger, K., Ledlow., M., \& Klypin, A. 1994, ApJL, 427, 87

\bibitem{4}
Conselice, C.J., \& Gallagher, J.S. 1998, MNRAS, 297, L34

\bibitem{5}
Dressler, A. 1980, ApJ, 236, 351

\bibitem{6} Dubinski, J., Mihos, J.C., \& Hernquist, L. 1996, ApJ, 462, 576

\bibitem{7} 
Frei, Z., \& Gunn, J.E. 1994, AJ, 108, 1476

\bibitem{8}
Gregg, M.D., \& West, M.J. 1998, Nature, 396, 549

\bibitem{9} 
Hernquist, L. 1990, ApJ, 356, 359

\bibitem{10}
Hickson, P. 1997, ARAA, 35, 357

\bibitem{11}
Hughes, J.P. 1998, in Untangling Coma Berenices: 
A New Vision of an Old Cluster, eds. A. Mazure, F. Casoli, F. Durret, \&
D. Gerbal (World Scientific: Singapore), 175

\bibitem{12}
Kormendy, J., Bender, R., Evans, A.S., \& Richstone, D. 1998, AJ, 115, 1823


\bibitem{14} Merritt, D. 1984, ApJ, 26, 37

\bibitem{15} Moore, B., Katz, N., Lake, G., Dressler, A., \& Oemler, Jr. A.
1996, Nature, 379, 613

\bibitem{16}
Sarazin, G. 1986, Rev. Mod. Phys., 58, 1

\bibitem{2} 
Soltan, A., \& Fabricant, D.G. 1990, ApJ, 364, 433

\bibitem{17}
Vikhlinin, A., Forman, W., \& Jones, C. 1994, ApJ, 435, 162

\end{thebibliography}
\end{document}